# "TRAVELLING WAVE" SOLUTIONS OF FITZHUGH MODEL WITH CROSS-DIFFUSION

F. Berezovskaya[1*], E. Camacho[2], S. Wirkus[3], G. Karev[4]

[1]Department of Mathematics, Howard University, Washington D.C., 20059,
fsberezo@hotmail.com
[2] Department of Mathematics, Loyola Marymount University, 90045
[3] Department of Mathematics & Statistics, California State Polytechnic University, Pomona, 91768
[4] Oak Ridge Institute for Science and Education, National Institute of Health, Bethesda, 20894

**Abstract**

The Fitzhugh-Nagumo equations have been used as a caricature of the Hodgkin-Huxley equations of neuron firing to better understand the essential dynamics of the interaction of the membrane potential and the restoring force and to capture, qualitatively, the general properties of an excitable membrane. Even though its simplicity allows very valuable insight to be gained, the accuracy of reproducing real experimental results is limited.

In this paper, we utilize a modified version of the Fitzhugh-Nagumo equations to model the spatial propagation of neuron firing; we assume that this propagation is (at least, partially) caused by the cross-diffusion connection between the potential and recovery variables. We show that the cross-diffusion version of the model, besides giving rise to the typical fast traveling wave solution exhibited in the original "diffusion" Fitzhugh-Nagumo equations, also gives rise to a slow traveling wave solution. We analyze all possible traveling wave solutions of the Fitzhugh-Nagumo equations with this cross-diffusion term and show that there exists a threshold of the cross-diffusion coefficient (the maximum value for a given speed of propagation), which bounds the area where "normal" impulse propagation is possible.

AMS classification scheme numbers: 34C20, 34C23, 92C99

## 1. Introduction

Hodgkin, Huxley, and Katz in the 1940's explored mathematically and experimentally the nature of nerve impulses. Their work revealed that the electrical pulses across the membrane arise from the uneven distribution between the intracellular fluid and the extracellular fluid of potassium ($K^+$), sodium ($Na^+$) and protein anions (Sherwood 2001). When a neuron is not sending a signal, it is said to be *at rest* (and at approximately -70 mV) and the inside of a neuron is more negative relative to the outside. The change in the $Na^+$ and $K^+$ permeability allows for the movement of ions in and out of the cell by means of opening and closing of ion channels. The influx of $Na^+$ and efflux of $K^+$ results in electrical potential difference. Triggered by a stimulus the $Na^+$ channels open, the influx of $Na^+$ ions increases, the membrane depolarizes, and the potential voltage reaches a *threshold* level typically between -50 and -55 mV. At this time an explosive depolarization takes place, which rapidly moves the potential to a maximum of +30 mV (Sherwood (2001)). At a similar rate, the $Na^+$ and $K^+$ channels close and open, respectively, initiating membrane repolarization caused by an efflux of $K^+$ ions. The potential drops back to *resting potential*. The intensity and magnitude of the forces behind repolarization are such that the membrane goes through a transient hyperpolarization of -10 mV below resting potential. This entire process of rapid change in potential from threshold to peak reversal and then back to the resting potential level is called *action potential,* impulse, or spike (see schematic diagram in Fig. 1).

This process by which a neuron fires was mathematically investigated by Hodgkin and Huxley, in 1952, with a four-variable model. In an effort to construct a simpler mathematical model of an excitable membrane FitzHugh (1961) proposed a two variable model. This model made it possible to illustrate the various physiological states involved in an action potential (such as resting, active, refractory, enhanced, and depressed) in the phase plane. One such portrait, which shows the spike-like behavior of the model, is given in Fig.2. The FitzHugh system, although a caricature of the Hodgkin-Huxley four equations, captures much of the same dynamical behavior.

A more realistic model is one that depends on both space and time since electric currents cross the membrane of the cell and move along its axon lengthwise inside and outside. This mechanism makes it possible for electrical signals to be transmitted over long distance and thus propagate throughout the membrane without ever weakening or decreasing their initial strength. With this in mind a mathematical model of the diffusion of current potential was first proposed and studied by FitzHugh (1961, 1969) and Nagumo et al. (1962). Recent models have been proposed where the spatial solutions are conditioned by the effects of cross-diffusion "control" or "interactions" between components of the system (see, for example, Kuznetsov et.al. (1996), Othmer and Stevens (1997), Berezovskaya and Karev (2000), etc.).

Motivated by these works we modified the FitzHugh model to include a cross-diffusion connection between the potential and recovery variables. We suppose that, due to the semiconductor nature of the nerve membrane, the cross-diffusion regulation plays an important (perhaps, crucial) role in the spatial spreading of potential. This version of the model will provide an avenue for investigating successful propagation of an excitable neuron but also propagation failures, which are extremely important for many applications. For example, the influence of certain drugs or external chemicals affect the rate at which sodium channels close and the rate at which potassium channels open, thus altering the normal dynamics of a firing potential membrane. A study conducted by Gubitosi-Klug and Gross (1996) showed that certain metabolites in ethanol accelerate the release of potassium ions from the brain cells. The increased potassium efflux in turn makes it difficult for cells to absorb enough calcium and thus inhibits the release of neurotransmitters (Highfield, 1996). The changes in the release of potassium ions result in the changes in the recovery phase of the excitable membrane. We include this effect of a generic drug by incorporating a cross-diffusion term in the original FitzHugh model.

In this work we explore the changes of the characteristics of the spatial propagation of nerve impulses brought by changes in the velocity of propagation and intensity of the cross-diffusion regulation. In particular, we are interested in the conditions of "normal" neuron firing propagation and investigate its possible violations.

The paper is organized in the following manner. Section 2 contains a brief description of local neuron dynamics within the framework of the FitzHugh model and the bifurcation portrait of the model. A cross-diffusion modification of the FitzHugh model aimed at providing an explanation of spatial modes like "traveling waves" is contained in Section 3. We show that fast and slow traveling waves can appear with respect to parameter values and follow their dependence by bifurcation analysis of corresponding wave systems; we show also that the "traveling spike" appears only before some threshold of the cross-diffusion coefficient. Section 4 contains the discussion of obtained results. Proofs of the statements of section 3 are given in the Appendix.

## 2. The Fitzhugh equations as a local membrane model

The original Fitzhugh model (1961) describing dynamics of the physiological states of a nerve membrane contains membrane potential variable $P$ and recovery variable $Q$ (which plays the role of all other three variables in the Hodgkin-Huxley model, sodium activation/inactivation and potassium activation). The variable $P$ shares the properties of both the membrane potential and excitability and thus describes the dynamics of the *rising phase* of neuron firing. The variable $Q$ is responsible for accommodation and refractoriness and thus represents the dynamics of the falling phase of the action potential. The equations are given by

$$P_\tau = I - Q - P^3/3 + P, \tag{1}$$
$$Q_\tau = \rho(a + P - bQ),$$

where $I$, $\rho$, $a$ and $b$ are parameters of the system. The stimulating current is defined by the variable $I$; $I<0$ corresponds to a cathodal shock and $I>0$ corresponds to an anodal shock. FitzHugh (1961, 1969) described the types of qualitative dynamics of this system. With the change of variables and parameters

$$P \to P\sqrt{3}, Q \to Q\sqrt{3} + I, \ \tau = t/(\rho b), \ k_1 = 1/b, \ \varepsilon = \rho b, \ k_2 = (I - a/b)/\sqrt{3} \tag{2}$$

the model is reduced to the form [1]

$$\varepsilon P_t = -P^3 + P - Q \equiv F_1(P, Q), \tag{3}$$
$$Q_t = k_1 P - Q - k_2 \equiv F_2(P, Q),$$

which contains only three parameters: $\varepsilon > 0$, $k_1 > 0$ and $k_2$. The complete phase-parameter analysis of the FitzHugh model (1) was given in (Volokitin and Treskov 1994). Additionally, it was shown (Khibnik et. al 1998) that a bifurcation of co-dimension 4 with symmetry "3-multiple neutral singular point with the degeneration" is realized in the vector field defined by system (3) in a vicinity of the parameter point $M$ ($k_1=1$, $k_2=0$, $\varepsilon=1$). The main results of these works are summarized in this section below.

The system (3) has from one (a non-saddle, i.e., a node or a spiral) up to three (two non-saddles and a saddle) singular points ($P^*, Q^*$) where $P^*, Q^*$ are common roots of $F_1(P,Q)$, $F_2(P,Q)$. A 3-multiple singular point $O(0,0)$ arises at $k_1=1$, $k_2=0$; it is a degenerated spiral sink if $\varepsilon \neq 1$ (see Berezovskaya and Medvedeva (1984)).

Let us consider the phase-parametric portrait (the bifurcation diagram) of the vector field (3), i.e., the partition of a vicinity of the parametric point **M** into all possible domains with topologically different phase portraits in a neighborhood of the point O (see Arnold (1993), Kuznetsov (1997)). For the sake of brevity, we call a cycle "small" if it contains a unique singular point, and "large" if it contains three singular points.

**Theorem 1.**

(*i*) *There exist a neighborhood of the parametric point **M** in which the cut of the bifurcation diagram of system* (3) *to the plane* ($k_1$, $k_2$) *is topologically equivalent to the diagram presented in Fig.-s 3a, 4 for arbitrary fixed $0<\varepsilon<1$ and to the diagram presented in Fig.-s 3b, 4 for arbitrary fixed $\varepsilon>1$. The boundary surfaces* (*lines at the* ($k_1$, $k_2$)-*cut at Fig.3*) *in the parameter space correspond to the following bifurcations:*

$S^\pm$: *appearance/disappearance of a pair of singular points on the phase plane;*

$H^\pm$: *change of stability of each of the non-saddle singular points in Andronof-Hopf supercritical bifurcation;*

*C: appearance/disappearance of a pair of limit cycles;*

$P^+$, $P^-$: *appearance/disappearance of a small limit cycle in one of two homoclinics of the saddle point containing a single non-saddle inside*[2];

$R^+$, $R^-$: *appearance/disappearance of a big limit cycle in one of two homoclinics of the saddle point containing two non-saddles inside*[3];

(*ii*) *Boundaries corresponding to the local bifurcations of the singular points in the parametric space $\{k_1,k_2,\varepsilon\}$ are described by the following equations:*

(1) *Surfaces*

$S^\pm$: $k_2 = \pm 2\sqrt{((1-k_1)^3/27)}$, $k_1 \leq 1$,

$H^\pm$: $k_2 = \pm(2+\varepsilon-3k_1)\sqrt{((1-\varepsilon)/27)}$, *under conditions that* $0<\varepsilon\leq 1$;

(2) *Lines*

SS: $k_1 = 1$, $k_2 = 0$,

---

[1] Volokitin and Treskov (1994) used the equivalent (for $b \neq 0$) changing of variables: $P \to -P\sqrt{3}$, $Q \to Q\sqrt{3} - 1$, $ak_1 = -(k_2\sqrt{3}+1)$, $\varepsilon k_1 = \rho$, $t = \varepsilon\tau$, $bk_1 = 1$.
[2] see right part of Fig.8 a,b
[3] see right part of Fig.8 c

$$SH^{\pm}: k_1 = \varepsilon, k_2 = \pm\ 2\sqrt{((1-\varepsilon)^3/27)},\ 0<\varepsilon\leq 1,$$

$$CH^{\pm}: k_1 = 2-\varepsilon, k_2 = \pm 4(1-\varepsilon)\sqrt{((1-\varepsilon)/27)},\ 0<\varepsilon\leq 1.$$

*Remark:* Surfaces (1) correspond to bifurcations of co-dimension 1; lines (2) correspond to bifurcations of co-dimension 2. The curves of intersections of $S^+$, $H^-$ and $S^-$, $H^+$ as well as the curve of intersection of $H^+$ and $H^-$ correspond to bifurcations of co-dimension "1+1".

The parameter space is divided into 21 domains of topologically different phase portraits. The parameter portraits possess certain symmetry. Due to this fact, the phase portraits are given by number and index *a,* but the respective *symmetric* phase portraits have no number in the parameter portrait.

Let us emphasize that in the framework of the FitzHugh model the spike-regime (see Fig. 1) is the trajectory $P(t)$ corresponding to the big separatrix loop in the phase plane. (Recall, that a big separatrix loop contains two singular points inside, while a small separatrix loop contains only one singular point inside; see footnotes[2,3].) The big separatrix loop is realized with parameter values $0<k_1<1$, $k_2$, $0<\varepsilon<1$ belonging to the boundaries $R^{\pm}$ of the parameter portrait. The dynamical regimes of the model similar to those shown in Fig.2 correspond to the phase curves in domains 4-6 and 10 of Figs. 3a,4.

## 3 Cross-diffusion model of a transmembrane potential

### 3.1 *Extending/Modifying Fitzhugh model*

In this section, we propose an extension of the FitzHugh spatial model to include the implicit (hypothetical) cross-diffusion mechanism of the spatial propagation of the firing process of the neuron. As mentioned previously, a cross-diffusion term involving the recovery variable could be used for modeling the effect of a generic drug that affects the flow of sodium and/or potassium.

We first give a very brief review of the well-known FitzHugh-Nagumo model (FHN-model). The simplest version of the "space-distributed" FHN-model (see, Nagumo *et al* (1962), FitzHugh (1969), and Chapter 6 of Murray (1993)) for transmembrane potential accounts for the "current" $W_1(t,x)$ along the axon due to the gradient of the potential in the point $x$, so that $W_1(t,x) \sim - P_x$. A more sophisticated approach takes into account another component of the current, $W_1(t,x) \sim - Q_x$, which defines the current against the gradient of the recovery variable. The total current $W(t,x)$ is then the sum of both components, and $P_t \sim -W_x$ (under zero local dynamics). Neglecting the possible currents of the recovery variable, we arrive at the model

$$\varepsilon P_t = -P^3 + P - Q + DQ_{xx} + \sigma P_{xx}, \qquad (4)$$

$$Q_t = k_1 P - Q - k_2,$$

where $t$ is time, $x$ is a one-dimensional space variable and non-negative constants $D$, $\sigma$ are the cross-diffusion and diffusion coefficients, respectively. The "diffusion" version of system (4), which corresponds to the case $D=0$, $\sigma>0$ is known as "*model FitzHugh-Nagumo*" (Nagumo et.al. (1964), FitzHugh (1969)). Many works were devoted to the study of its dynamics, and in particular, to the investigation of "traveling wave" solutions (see, e.g., Hastings (1976), Evans *et al.* (1982), etc.). A bifurcation approach applied to the study of traveling impulses and trains (see Kuznetsov (1997)) allowed to reveal "fast" and "slow" waves that can exist with the same values of "local parameters" $\varepsilon$, $k_1$.

To make clearer the role of the spatial distribution of the recovery variable (along the axon) and the meaning of the cross-diffusion term in the impulse propagation, we consider a cross-diffusion version of system (3):

$$\varepsilon P_t = -P^3 + P - Q + DQ_{xx} \equiv F_1(P, Q) + DQ_{xx}, \qquad (5)$$

$$Q_t = k_1 P - Q - k_2 \equiv F_2(P, Q).$$

Mathematically, cross-diffusion equations possess special properties, which facilitate their research (Wei-Ming Ni, 1998, Berezovskaya, 1998). In what follows, we explore "traveling wave" solutions of system (5):

$$P(x,t) = P(x + Ct) \equiv p(\xi), \quad Q(x,t) = Q(x + Ct) \equiv q(\xi)$$

where $\xi = x + Ct$ and positive $C$ is the velocity of the wave propagation. It can be checked that $(p(\xi), q(\xi))$ satisfy the following *two-dimensional "wave system"*:

$$(\varepsilon C^2 - Dk_1)/C \, p_\xi = F_1(p,q) - DF_2(p,q)/C^2, \qquad (6)$$

$$Cq_\xi = F_2(p,q).$$

For convenience, let us denote $\alpha = C^2/(\varepsilon C^2 - Dk_1)$ and change the independent variable $\eta = \xi/C$, then the wave system (6) becomes

$$p_\eta = \alpha \, (F_1(p,q) - DF_2(p,q)/C^2), \qquad (7\beta)$$

$$q_\eta = F_2(p,q).$$

where $\beta = \text{sign}(\alpha)$. This system is defined for $\varepsilon C^2 \neq Dk_1$.

Further, we study phase behaviors of (7) dependending on "local" parameters $\varepsilon$, $k_1$, $k_2$ for arbitrary fixed values of constants $D$ and $C$ such that $C^2 \neq Dk_1/\varepsilon$. In other words, the domain of the parameters *D, C* under fixed values of $\varepsilon, k_1, k_2$ is divided into two domains, in which the system exhibits qualitatively different behavior, by the parabola $C^2=Dk_1/\varepsilon$.

### 3.2. Wave system of the cross-diffusion Fitzhugh model

System (7+) with *C, D* such that $C^2>Dk_1/\varepsilon$ is called *fast* wave system, whereas system (7−) with $C^2<Dk_1/\varepsilon$ is called *slow* wave system of model (5). For both cases system (7β), evidently, has from one up to three singular points (*p\*,q\**) whose coordinates satisfy equations:

$$F_1(p,q)=0, \quad F_2(p,q)=0.$$

Two singular points coincide and form two-multiple point with the parameter values belonging to boundaries $S^\pm$ (see Fig.3 and Theorem 1). Thus, both the fast and slow wave systems have three singularities inside the curvilinear angle that is formed by $S^\pm$, and a single singular point outside the angle. Note that this unique singular point is a non-saddle (a node or spiral) for the fast wave system whereas it is a saddle for the slow one. Inside the parameter angle the fast wave system has two non-saddles and a saddle whereas the slow one has two saddles and a non-saddle (see Mathematical Appendix, Proposition 3). At $k_1=1$, $k_2=0$ the points coincide in three-multiple singular point O(0,0) which is a non-hyperbolic spiral if $\varepsilon \neq 1$ for the fast wave system and a non-hyperbolic saddle for the slow wave system.

**Theorem 2.**

(i) *Let $C^2>Dk_1/\varepsilon$. There exist a neighborhood of the parameter point $M(k_1=1, k_2=0, \varepsilon=1)$ in which the vector field defined by system (7+) in a neighborhood of the phase point (p=0, q=0) has a bifurcation diagram, whose cut to the plane $(k_1, k_2)$ is topologically equivalent to the one presented in Fig.3a, 4 for arbitrary fixed $0<\varepsilon<1$ and in Fig.3b, 4 for arbitrary fixed $\varepsilon>1$.*

*The boundaries in $(k_1, k_2, \varepsilon)$- parameter space (lines at the $\varepsilon$-cut at Fig.3) correspond to the same bifurcations that have been mentioned in Theorem 1 and have the same equations for $S^\pm$, $H^\pm$, SS and $SH^\pm$.*

(ii) *Let $0<C^2<Dk_1/\varepsilon$. There exist a neighborhood of the parameter point $M(k_1=1, k_2=0, \varepsilon=1)$ in which the vector field defined by system (7-) in a neighborhood of the phase* *point ($p=0,q=0$) has a bifurcation diagram, whose cut to the plane ($k_1$, $k_2$) is topologically equivalent to the one presented in Fig.5a, 6 for arbitrary fixed positive $\varepsilon<1$ and in Fig.5a, 6 for arbitrary fixed $\varepsilon>1$.*

The boundary surfaces in the parameter space correspond to the following bifurcations:

$S^{\pm}$: *appearance/disappearance of a pair of the singular points;*

$H_{\pm}$: *change of stability of the unique non-saddle singular point in Andronof-Hopf supercritical bifurcation;*

$P_+$, $P_-$: *appearance/disappearance of a small limit cycle in non-local homoclinic bifurcations of the saddle point$^2$;*

$L_+$, $L_-$: *upper and lower* (*respectively*) *heteroclinics of saddle singular points$^4$ .*

Equations of the parameter boundaries corresponding to the local bifurcations are the same that those given in Theorem 1 *for $S^{\pm}$, SS and $SH_{\pm}=SH^{\pm}$*.

The parameter space of system (7-) is divided into 10 domains of topologically different phase portraits. The parameter portraits possess certain symmetries. Due to this fact, the phase portraits are given by number and index *a,* but the respective *symmetric* phase portraits have no number in the parameter portrait.

*Remark*. The proofs of Theorems 1 and 2 (given in the Appendix) contain the analytical description of local bifurcations of the systems. Non-local bifurcations were investigated mostly by computation.

### 3.3. *Traveling wave solutions of PDE and their profiles as solutions of wave ODE*

The correspondence between traveling wave solutions (6) of model (5) and orbits of its wave system (7) is schematically given in Figures 7 through 9 (see, e.g., Volpert *et al* (1994), Berezovskaya and Karev (1999, 2000))**.**

Heteroclinic, homoclinic orbits, and limit cycles of a wave system correspond to wave fronts, pulses, and train solutions of the model, respectively. A heteroclinic curve connects two singular points; a homoclinic curve connects a single singular point to itself; and a phase limit cycle encloses an odd number of singular points.

---

[4] *see middle part of Fig.7a,b*

Thus, the description of all possible wave solutions of PDE system (5) is reduced to the analysis of phase curves and bifurcations in the wave ODE system (7) in which *C* is an "additional" parameter. We make this correspondence more formal with the following statement.

**Proposition 1.**

*i*) *A spatially homogeneous solution u= u\* of the model corresponds to a singular point (u,v) =(u\*, 0) of the vector field defined by the wave system;*

*ii*) *a wave front of the model corresponds to a heteroclinic curve of the wave system which joins singular points with different u-coordinates, see Fig.*7**a,b***;*

*iii*) *a wave impulse corresponds to a homoclinic curve (separatrix loop) of a singular point of the wave system (see Fig.8 where small loops (a,b) and big loop (c) are shown);*

*iv*) *a wave train of the model corresponds to a limit cycle in the (u,v) phase plane of the wave system, see Fig.9.*

**3.4.** *Fast and slow traveling waves of the cross-diffusion system*

We say that model (5) has a traveling wave solution of the given type in some parametric domain if for any parametric point from this domain there exists an initial value $(p_0, q_0)$ for the wave system (7$\beta$) such that the trajectory of the wave system corresponds to the traveling wave solution of (5) of the given type. Applying Proposition 1 and Theorem 2 we can now describe the fast and slow traveling wave solutions of model (5).

**Theorem 3.** *Model (5) has the fast traveling wave solutions (i.e., with $C^2 > Dk_1/\varepsilon$) of the following types:*

*the fronts in every domain of the portrait Fig. 3a except the domain 1;*

*the trains in domains 3a-b, 5a-b, 6a-b, 7a-b, 8-11, 12a-b, 13a-b, 14a-b; additionally, the model has two trains, differing in their "amplitudes", in domains 5a-b, 8, 12a-b, 14a-b, and three different trains in the areas 11 and 13a-b (see Fig.4);*

*the impulses for the parameter points $(k_1, k_2, \varepsilon)$ belonging to the boundaries $P^+$, $P^-$ and $R^+$, $R^-$, see Fig.8a,b,c.*

**Theorem 4.** *Model (5) has the slow traveling wave solutions (i.e., with $C^2 < Dk_1/\varepsilon$) of the following types:*

*the fronts in every domain of the portrait except 1`; monotonous fronts with the maximal "amplitude" for the points $(k_1, k_2, \varepsilon)$ belonging to the boundaries $L_+$, $L_-$;*

*the trains in the domains 4`a, 5` (see Fig.6);*

*the impulses for the points $(k_1, k_2, \varepsilon)$ belonging to the boundaries $P_+$, $P_-$ (see Fig 8).*

### 3.5. *Stationary distribution of the model*

The stationary spatial distribution of model (5) is defined by the equations $P_t = Q_t = 0$. Then system (5) takes the form:

$$DQ_{xx} = -F_1(P,Q), \qquad (8)$$
$$F_2(P,Q) = 0$$

and can be easily reduced to the second order equation with respect to $P$ and $x$:

$$Dk_1 P_{xx} = -(k_2 + P(1-k_1) - P^3). \qquad (9)$$

Note that the wave system of model (5):

$$\varepsilon C p_\xi = F_1(p, q) + D q_{\xi\xi},$$
$$C Q_\xi = F_2(p, q)$$

coincides with system (8) if $C=0$. Traveling wave solution with $C=0$ is called a *standing wave*. One could write (9) as the Hamiltonian system

$$P_x = W, \qquad (10)$$
$$W_x = (P^3 - P(1-k_1) - k_2)/(Dk_1),$$

whose Hamiltonian is of the form:

$$H(P,W) = (2Dk_1 W^2 + 2P^2(1-k_1) + 4Pk_2 - P^4)/(2Dk_1) + \text{const.} \qquad (11)$$

System (10) has from one (a center) up to three (two saddles and a center) equilibria $(P^*, 0)$, where $P^*$ satisfies the equation

$$P^3 - P(1-k_1) - k_2 = 0.$$

The parameter portrait of (10) presented in Fig.10a is the "light version" of the portrait in Fig.5 (see, for example, Kuznetsov, Antonovsky et. al, 1991). This portrait contains boundaries $S^{\pm}$, whose equations were given in Theorem 1, and line $L$: $k_2 = 0$. On the boundary $L$ with any $0 < k_1 < 1$ the phase portrait of the model contains the upper and lower heteroclinics (see Fig. 10c); if the parameter values lie inside curvilinear angles $LS^+$ or $LS^-$ then the phase portraits have left or right separatrix loop, respectively (see Fig. 10 a,b).

**Proposition 2.** *Depending on parameters $k_1$, $k_2$ (see parameter portrait in Fig.10, left) three types of standing waves are realized in model (5) for suitable initial values $(p,q)$:*

i) trains corresponding to limit cycles (*Fig.10a,b,c*);

ii) small impulses corresponding to small homoclinics (*Fig.*10*a,b*) and

iii) fronts corresponding to heteroclinics (*Fig.* 10*c*)

## 4. Discussion and conclusion

### 4.1. *Comparing of bifurcation diagrams*

We have shown that the parameter portrait of the local FH-model (3) as well as the fast wave systems (7+) presented in Fig.3, has many features in common with the parameter portrait of the slow wave systems (7-) shown in Fig.5. For example, the boundaries $S^{\pm}$ are congruent; the boundaries $H^{\pm}$ of the Hopf bifurcation of the former system continue smoothly the boundaries $H_{\pm}$ of the Hopf bifurcation of the latter system and belong to the common curve of neutrality; the boundaries $P^+$ and $P^-$ of the homoclinic bifurcation of the first system continue smoothly corresponding boundaries $P_+$ and $P_-$ of the other system, etc. (see Appendix for details). As a result, one could easily compare the qualitative behaviors of both systems.

We emphasize that the parameter portrait in Fig.3 of the local FH-model also corresponds to the wave system of the cross-diffusion model with a small cross-diffusion coefficient $D$ (under fixed propagation speed $C$) while the parametric portrait in Fig.5 corresponds to the same model with "large" cross-diffusion coefficient $D$. Hence, these portraits describe the model behavior *before* and *after* the threshold $D= \varepsilon C^2/k_1$ accordingly.

### 4.2. *Scenarios of appearance and transformations of the traveling waves*

The problem of our interest is the appearance and transformations of the traveling wave solutions depending on the model parameters $D$ and $C$ that characterize the axon abilities for the firing propagation. One could assume that these characteristics may change as a result of influence of certain drugs or external chemicals. We now trace the transformation of the traveling wave solutions by varying these parameters under the supposition that parameters $k_1$, $k_2$, $\varepsilon$ have arbitrary fixed values close to $k_1=1$, $k_2=0$, $\varepsilon=1$. Let the (positive) value of the speed propagation $C$ be fixed and suppose the positive cross-diffusion coefficient increases. For $D=0$ the wave system of the model coincides with the local Fitzhugh model. This model demonstrates a spike (shown in Fig.1) if $(k_1^*, k_2^*, \varepsilon^*)$ belongs to the boundary $R^+$. The wave system describes "pseudo-waves" and, in reality, there is no firing propagation.

Let $D>0$. Due to Theorems 3, 4 if $D< k_1 C^2/\varepsilon$, the model (at $(k_1^*, k_2^*, \varepsilon^*)$ belonging to the boundary $R^+$) has a traveling spike spreading along the axon with velocity $C$. A velocity $C$ of the spike propagation must be greater that $\sqrt{Dk_1/\varepsilon}$ and the amplitude of the spike $\{p(\xi), q(\xi)\}$ is "large", i.e. greater than $[p_3-p_1, q_3-q_1]$, where $p_1<p_2<p_3$ are the roots of the polynomial $F_1(p, k_1p - k_2) = -p^3 + p(1-k_1) + k_2$ and $q_i = q(p_i) = k_1 p_i - k_2$, $i=1,2,3$.

Let us note that as $C\to\infty$ parameter $\alpha(C) \to 1/\varepsilon$, hence the wave system (7) formally becomes the local system (3). The bifurcation diagram given in Figs. 3, 4 allows us to identify all other possible waves, namely, trains with "large" and "small" amplitude (i.e., more or less, respectively, than either $[p_2-p_1, q_2-q_1]$ or $[p_3-p_2, q_3-q_2]$) and fronts with non-monotonic tails.

On the contrary, if $D> \varepsilon/k_1 C^2$ then a traveling spike does not exist. The only possible traveling waves are "small" trains, impulses or fronts with the amplitudes less than $[p_3-p_1, q_3-q_1]$. The velocity $C^*$ of propagation of these waves is less than $\sqrt{Dk_1/\varepsilon}$. In particular, there exist standing waves ($C^*=0$), which look like oscillations, "small" impulses, or wave fronts in space depending on the model parameters (see Fig. 10). Note that the bifurcation diagram given in Figs. 5, 6 allows us to describe transformations of waves with the changing of their velocity. If a parametric point $(k_1^*, k_2^*, \varepsilon^*)$ belongs to domain 6a` then the sole traveling wave solution $\{(p(\xi), q(\xi)\}$ is the wave front with non-monotonic tail. When the velocity $C$ increases (under the condition $C<\sqrt{Dk^*_1/\varepsilon^*}$), the system intersects the boundary $L^+$ and enters into domain 7`. There exist two slow wave-fronts with non-monotonic tails moving with the same velocity from the right to the left; their amplitudes are $[p_3-p_2, q_3-q_2]$ and $[p_2-p_1, q_2-q_1]$ correspondingly. When $C$ increases, both waves becomes monotonic; for $C=\sqrt{Dk^*_1/\varepsilon^*}$ the first equation of wave system (7-) is degenerate and describes the curve $q = -p^3 + p(1-k_1) + k_2$ that smoothly joins the points $p_1, p_2, p_3$. Further increase of $C$ leads to a "transformation" of the slow wave system into the fast one, and thus to the appearance of waves similar to the spike spreading along an axon. A behavior of the model under critical values $D= k_1 C^2/\varepsilon$, evidently, cannot be studied in the framework of the two-dimensional model (5).

**4.3. *Conclusion***

We utilized a modified version of the Fitzhugh-Nagumo equations to model the spatial propagation of neuron firing; we assumed that this propagation is essentially caused by the cross-diffusion connection between the potential and recovery variables. This modification, which includes the implicit (although hypothetical) cross-diffusion mechanism could help to explore the effect of a generic drug in the neuron firing process. The incorporation of the cross-diffusion mechanism is motivated by the experimental observation that certain metabolites in ethanol accelerate the release of potassium ions and thus have a direct affect on the flow of potassium or sodium (Gubitosi-Klug and Gross 1996). Since the ionic concentration of potassium determines how quickly hyper-polarization occurs, altering the flow of potassium will change the dynamics of the recovery process. In addition, at any given time "there are circulating currents that cross the membrane and flow lengthwise inside and outside the axon and the membrane current and potential vary with distance as well as with time" (FitzHugh 1969). A generic drug that alters the flow of potassium has an effect on the neuron returning to its rest potential therefore it is natural to incorporate a spatial component in the membrane potential equation involving the recovery variable.

The mathematical problem of our interest was the appearance and transformations of the traveling wave solutions, which depended on the model parameters *D* (the cross-diffusion coefficient) and *C* (the propagation speed) that characterize the axons abilities for the firing propagation. We studied the wave system of the cross-diffusion version of the model and explored its bifurcation diagram. We studied the wave system of the cross-diffusion version of the model and explored its bifurcation diagram.

We have shown that the cross-diffusion model possesses a large set of traveling wave solutions; besides giving rise to the typical "fast" traveling wave solution exhibited in the original "diffusion" Fitzhugh-Nagumo equations, it also gives rise to a "slow" traveling wave solution. A more sophisticated approach showed that instead of a "one-parametric" set of waves ordered by the propagation speed *C,* one should consider a two-parametric set of traveling wave solutions with parameters (*C,D*). We then proved that in the parametric space (*D,C*) (under fixed values of other model parameters $\varepsilon$, $k_1$, $k_2$) there exists a parabolic boundary, $D^* = KC^2$, where constant $K = \varepsilon/k_1$, which separates the domains of existence of the fast and slow waves. The system behavior qualitatively changes with the intersection of this boundary. Let us emphasize that the "traveling spike" that we consider as the "normal" propagation of a nerve impulse is a "fast" traveling wave. On the other hand, the domains of fast (with $C^2 > Dk_1/\varepsilon$) and slow (with $C^2 < Dk_1/\varepsilon$) waves are evidently the same as the areas with small ($D < KC^2$) and large ($D > KC^2$) values of the cross-diffusion coefficient, respectively. Hence, the parabola $D^* = KC^2$ bounds the area where the "normal" spike propagation is possible. After the intersection of this boundary, due to very large of the cross-diffusion coefficient or too small speed of impulse propagation, a "normal" propagation of the nerve impulse is impossible and some violations are inevitable: nerve impulses propagate with decreasing amplitude or as damping oscillations.

So, the cross-diffusion regulations in the FitzHugh model allowed us to observe the propagation of spikes and spike-like oscillations but restricted their velocities from below or, equivalently, maintained the upper boundary for the cross-diffusion coefficient. It means that if, by any reasons (e.g., as a result of the effect of a generic drug) the speed of transmission of a signal along the axon is reduced, then the "normal" neuron firing propagation in the form of a traveling spike is impossible. The increase of the cross-diffusion coefficient beyond the "normal" value implied the same result.

## 5. Mathematical Appendix. Proof of Theorems

### 5.1. *Plan of the proof*

The proof consists of two steps. Firstly, by the non-degenerate changing of variables we reduce the initial model (3) to the generalized Lienard form. The same transformation applied to the wave system (7β) also reduces it to the generalized Lienard form. These Lienard form systems (the former with $t$ and the latter with $\xi$ as independent variables) have vector fields possessing many common characteristics. We remark that this transformation reduces the cross-diffusion modification (5) of system (3) to the cross-diffusion modification of the initial model (3) transformed to the generalized Lienard form.

Secondly, we carry out the bifurcation analysis of the resulting vector fields. This analysis is facilitated by to the polynomial Lienard form of the systems in which many local bifurcation are represented in their *normal forms* (see Arnold 1983, Dumortier *et al* 1991, etc.). Using suitable bifurcation diagrams of the normal form system one can obtain the phase-parameter structure of the initial model (3) and wave system (7β). Note that the wave system (7β) can describe different kinds of bifurcations with respect to $\beta$, that is the sign of $\alpha(C)$.

## 5.2. Lienard form of the wave system

By the change of variables $(P,Q) \to (U, Z)$:

$$U = Q + k_2, \quad Z = F_2(P,Q) \equiv k_1 P - Q - k_2 \tag{12}$$

the local model (3) is transformed to the generalized Lienard form:

$$U_t = Z, \tag{13}$$

$$\varepsilon Z_t = f(U) + Z(g_1(U) + ZG(U,Z)) \equiv \Phi(U,Z),$$

where

$$f(u) = -u^3/k_1^2 + u(1-k_1) + k_1 k_2 \tag{14}$$

$$g_1(u) = (1 - \varepsilon) - 3u^2/k_1^2,$$

$$G(u,z) = -(3u + z)/k_1^2.$$

Note, that model (5) after transformation (12) reads as the cross-diffusion modification of (13) with the coefficient $Dk_1$:

$$U_t = Z, \tag{15}$$

$$\varepsilon Z_t = \Phi(U,Z) + Dk_1 U_{xx},$$

A *traveling wave solution* of system (14-15) is defined as a pair of bounded functions

$$U(x,t) = u(x + Ct) \equiv u(\xi), \quad Z(x,t) = z(x + Ct) \equiv z(\xi) \tag{16}$$

where $C>0$ is a velocity of propagation. One can verify that after introducing an independent variable $\eta = \xi/C$ the functions $\{u(\eta), z(\eta)\}$ satisfy the *wave system*

$$u_\eta = z, \tag{17\beta}$$

$$z_\eta = \alpha\Phi(u,z) \equiv F(u,z;\alpha)$$

Here $\alpha = C^2/(\varepsilon C^2 - Dk_1)$ if $C^2 \neq Dk_1/\varepsilon$, $\beta = \text{sign}(\alpha)$ and $F(u,z;\alpha) = \alpha f(u) + \alpha z g_1(u) + \alpha z^2 G(u,z)$ with $f(u)$, $g_1(u)$ and $G(u,z)$ given by (14). Note, that $\alpha = 1/\varepsilon$ if $D=0$ and $\varepsilon \neq 0$. It means that the vector field defined by system (17+) coincides with the vector field defined by system (13).

Let's now replace the capital letters in (12) by small letters, reduce $p$ and $q$ via

$$p = (z+u)/k_1, \quad q = u - k_2 \quad (k_1 \neq 0),$$

and substitute into system (7β). Dividing equation for $u_\xi$ by $C \neq 0$ we get system (17β).

## 5.3. Main characteristics of the vector fields depending on α

Consider the generalized Lienard vector field:

$$\mathbf{J} = z\,\partial/\partial u + F(u,z;\alpha)\,\partial/\partial z \tag{A1}$$

where

$$F(u,z;\alpha) = \alpha(f(u) + z(g_1(u) + zG(u,z))), \tag{A2}$$

$$f(u) = -a\,u^3 + u\delta_1 + \delta_2,$$

$$g_1(u) = \delta_3 - 3au^2,$$

$$G(u,z) = -a(3u + z)$$

with positive constant $a$, arbitrary $\alpha \neq 0$ and "small" parameters $\delta_1$, $\delta_2$, $\delta_3$.

**Proposition 2.** *Let $\delta_1 = 1-k_1$, $\delta_2 = k_2 k_1$, $\delta_3 = 1-\varepsilon$, $a = 1/k_1^2$ and $\alpha* = C^2/(\varepsilon C^2 - Dk_1)$. Then vector field (A1, A2) coincides with vector field given by system (13) if $\alpha = 1/\varepsilon$ and with vector field given by system (17$\beta$) if $\alpha = \alpha*$.*

For any $\alpha$, $0 < \alpha < \infty$ vector field (A1, A2) has at least one singular point $(u_0, 0)$ where $u_0$ is a root of the cubic polynomial:

$$f(u) = -au^3 + u\delta_1 + \delta_2.$$

**Proposition 3.** *For positive $\alpha$ vector field (A1), (A2) has a single (non-saddle) singular point outside the curvilinear angle formed by curves $\delta_2 = \pm 2(\delta_1^3/27a)^{1/2}$ and three (two non-saddles and one saddle) singular points inside this angle;*

*for negative $\alpha$ vector field (A1), (A2) has a single (saddle) or three (two saddles and one non-saddle) singular points outside and inside, respectively, the curvilinear angle.*

*Proof.* The Jacobian of (A1), (A2) at point $(u_0, 0)$ is: $A(J) = \begin{pmatrix} 0 & 1 \\ f_u(u_0) & g_1(u_0) \end{pmatrix}$.

For $\delta_1 = 0$ the vector field has the unique singularity $(u_0, 0) = ((\delta_2/a)^{1/3}, 0)$. The Jacobian determinant at $(u_0, 0)$, $\det(A) = 3\alpha u_0^2$, is a positive for positive $\alpha$, so the singular point is non-saddle, and a negative for negative $\alpha$, so the point is a saddle. Due to standard continuity arguments, for small $\delta_1$ the polynomial $f(u)$ has a root $u_0*$ close to $u_0$, $\det(A)$ preserves its sign and the type of singular point remains unchanged.

For arbitrary $\delta_1 \neq 0$ $f(u)$ may have two additional roots. An appearance of these roots corresponds to the appearance of two additional singular points of the vector field. Generally, these additional points are a saddle and a node. They are seen to come into existence when

$$-au^3 + u\delta_1 + \delta_2 = 0, \quad -3au^2 + \delta_1 = 0$$

and one can easily find the boundaries of the *curvilinear angle* in $(\delta_1, \delta_2)$-space that correspond to existence of two-multiple roots $u_0$ of polynomial $f(u)$.

### 5.4. Beginning of proofs of Theorems 1,2

The following statement is given in [Khibnik et al., 1998].

**Theorem KKH**. *There exists a neighborhood of the parametric point $(\mu_1, \mu_2, \mu_3)=(0,0,0)$ in which the system*

$$y_1' = y_2 \tag{NF}$$
$$y_2' = \mu_2 + \mu_1 y_1 - y_1^3 + y_2(\mu_3 - y_1^2) + O(\|y\|^3) \equiv G(y_1, y_2; \mu_1, \mu_2, \mu_3) + O(\|y\|^3)$$

*in a neighborhood of the phase point $(y_1=0, y_2=0)$ has a bifurcation diagram, whose cut to the plane $(\mu_1, \mu_2)$ for arbitrary $\mu_3 > 0$ is topologically equivalent to the diagram presented in Figs. 3, 4.*

Evidently, by scaling $y_1 \to a_1^{1/3} y_1$, $y_2 \to b y_2$, $\nu_1 = a_1^{1/3} \mu_1$, $\nu_2 = \mu_2$, $\nu_3 = b \mu_3$ ($a > 0$, $b > 0$) the function $G(y_1, y_2; \mu_1, \mu_2, \mu_3)$ of system (NF) transforms to the function $G_1(y_1, y_2; \nu_1, \nu_2, \nu_3) = \nu_2 + \nu_1 y_1 - a_1 y_1^3 + y_2(\nu_3 - B y_1^2)$ where $B = b a_1^{2/3}$ and $\nu_1, \nu_2, \nu_3$ are small parameters if $\mu_1, \mu_2, \mu_3$ were small.

Now statement (i) of Theorem 1 follows from Theorem KKH and Proposition 2 if we put $\alpha = 1/\varepsilon$, $a_1 = 1/\varepsilon k_1^2$, $b = 3/(\varepsilon k_1^2)^{1/3}$, $\varepsilon \nu_1 = (1 - k_1)$, $\varepsilon \nu_2 = k_1 k_2$, $\varepsilon \nu_3 = (1 - \varepsilon)$ with $\varepsilon > 0$, $k_1 \neq 0$ in (A1), (A2). Statement (i) of Theorem 2 follows from Theorem KKH and Proposition 2 if we put $\alpha = \alpha *$, $a_1 = (\alpha *)/k_1^2$, $b = 3((\alpha *)/k_1^2)^{1/3}$, $\nu_1 = (1 - k_1)\alpha *$, $\nu_2 = k_1 k_2 \alpha *$, $\nu_3 = (1 - \varepsilon)\alpha *$ for any fixed $C \neq 0$ and $D < \varepsilon C^2 / k_1$.

In order to prove statement (ii) of Theorem 2 as well as for describing bifurcation boundaries in the parameter portraits of Figs. 3 and 5, we consider below behaviors of vector field (A1),(A2) depending on parameters $\delta_1, \delta_2, \delta_3$ as well as on parameter $\alpha$.

Let $(u_0, 0)$ be a singular point of (A1), (A2), that is $u_0$ is a root of polynomial $f(u)$. For study of local bifurcations we shift $(u_0, 0)$ to the origin:

$$v = u - u_0 \Rightarrow u = v + u_0 \Rightarrow u' = v',$$

and get the system:

$$v' = z, \tag{A4}$$
$$z' = \alpha(a_{10} v + a_{01} z + a_{20} v^2 + a_{02} z^2 + a_{11} z v + a_{12} v z^2 + a_{21} v^2 z + a_{30} v^3 + a_{03} z^3).$$

Here coefficients $a_{ij}$ ($i,j = 1,2,3$) are:

$$a_{10} = \delta_1 - 3au_0^2, \quad a_{01} = \delta_3 - 3au_0^2, \tag{A5}$$
$$a_{20} = -3au_0, \quad a_{02} = -3au_0, \quad a_{11} = -6au_0,$$
$$a_{30} = -a, \quad a_{21} = -3a, \quad a_{12} = -3a, \quad a_{03} = -a.$$

A bifurcation of $(u_0, 0)$ in system (A1, A2) coincides with the bifurcation of $(0,0)$ in system (A4, A5) whose Lienard form allows respectively easy to formulate the bifurcation conditions (see, for example, Arnold, 1983, Guckenheimer & Holmes, 1983, Bautin & Leontovich, 1976).

The following statement holds (Volokitin, Treskov, 1994).

***Lemma A*1.** Singular point O of vector field (A4) is

1) two-multiple one, corresponding to codim 1 bifurcation of appearance / disappearance of two equilibria in the phase plane, if $a_{10}=0$, $a_{20} \neq 0$;

2) neutral spiral[5], corresponding to Andronov-Hopf bifurcation of codim 1 if $a_{01}=0$, $\alpha a_{10}<0$ and *the first Lyapunov value* $L_1 \equiv -\alpha^2\{a_{11}(-a_{20}+\alpha a_{10}a_{02}) + a_{10}(-3\alpha a_{10} a_{03} +a_{21})\} \neq 0$;

3) three-multiple one, corresponding to codim 2 bifurcation of appearance / disappearance of three phase singular points, if $a_{10}=0$, $a_{20}=0$ and $a_{30} \neq 0$;

4) two-multiple neutral[5] one, corresponding to Bogdanov-Takens bifurcation of codim 2, if $a_{01}=0$, $a_{10}=0$ and $a_{11}a_{20} \neq 0$;

5) neutral spiral with the degeneration, corresponding to the codim 2 bifurcation "Zero of the first Lyapunov value", if $a_{01}=0$, $\alpha a_{10}<0$, $L_1 =0$, and *the second Lyapunov value* [Bautin, Leontovich, 1976] $L_2 \equiv -\pi/24(-\alpha^4(a_{11}^3(a_{02} + a_{20})- \alpha^3 (15a_{20}^2 a_{03} -3a_{02}^2 a_{21} + a_{20} (9a_{11} a_{03} -5 a_{02} a_{21}) - a_{11}^2 (a_{21}+ a_{03})+4 a_{20} a_{11}(a_{30}+ a_{12}) - a_{02} a_{11} (7 a_{03} +5 a_{12} ) + 3 a_{12} (a_{21} +2a_{03} ) - 3\alpha^2 a_{30} a_{21}) \neq 0$.

**Corollary 1.** Singular point $(u_0,0)$ of vector field (A1, A2) where $u_0$ is a root of polynomial $f(u)= -au^3 + u\delta_1 + \delta_2$ ($a \neq 0$), is

1) two-multiple one in the phase plane, if $u_0^2 =\delta_1/(3a) \neq 0$ and $u_0 \neq 0$ and three-multiple one if $u_0=0$ and $\delta_1=\delta_2=0$;

2) two-multiple neutral one if $3au_0^2 = \delta_1 = \delta_3$ and $u_0 \neq 0$;

3) neutral spiral for $\alpha >0$ if $u_0^2 =\delta_3/(3a)$, $\delta_1 -3au_0^2 <0$ and $L_1 \equiv 3a\alpha^2(\delta_3 +\delta_1)(\alpha(\delta_3 - \delta_1)+1) \neq 0$; neutral spiral for $\alpha <0$ if $u_0^2 =\delta_3/(3a)$, $\delta_1 -3au_0^2 >0$, and $L_1 \neq 0$;

4) neutral spiral with $L_1=0$ for $\alpha >0$ if $u_0^2 =\delta_3/(3a)$, $\delta_1 -3au_0^2 <0$.

**Corollary 2.** For $\delta_3 <0$ the bifurcation diagram of vector field (A1, A2) contains only domains 1 and 2 if $\alpha >0$ (see Fig.-s 3,4 b) and domains 1`,2`,3`a, 3`b if $\alpha<0$ (see Figs. 5,6 b).

*Proof of Corollary* 1.

---

[5] neutral singular point has zero trace of Jacobian ; neutral spiral is a singular point having imaginary eighenvalues

Statements 1 and 2 of Corollary 1 can be checked easily. Let us prove only the third and fourth statements. The value $L_1$ can vanish only if $\alpha(\delta_3 - \delta_1) + 1 = 0$ or $\delta_3 + \delta_1 = 0$. Note that $\delta_1 < \delta_3$ if $\alpha > 0$ and $0 < \delta_3 < \delta_1$ if $\alpha < 0$ from conditions of spiral neutrality, thus, the first equality is not valid for $\alpha$ of any sign. Hence, $L_1 = 0 \Leftrightarrow \delta_3 = -\delta_1$, and so $L_1$ can vanish only for $\alpha > 0$. Note now that for these values $L_2 = -18a\alpha^2(3 + 50\ a\alpha\ u_0^2 + 72\ a^2\alpha^2\ u_0^4) = -6a^2\alpha^2(9 + 50\ \alpha\delta_3 + 24\ \alpha^2\ \delta_3^2) < 0$ for $\delta_3 > 0$. Statements 3) and 4) are proved.

*Proof of Corollary 2.* For any $\delta_3 < 0$ the divergence of vector field (A1, A2), $\mathrm{div} J = \alpha(g_1(u) + 2zG(u,z) + z^2 G_z(u,z)) = \alpha((\delta_3 - 3au^2) - 3az(2u+z))$, keeps its sign in some neighborhood of zero phase point containing inside all singular points of the field: $\mathrm{div} J < 0$ for $\alpha > 0$ and $\mathrm{div} J > 0$ for $\alpha < 0$. Thus, the vector field has no closed orbits in this neighborhood; any non-saddle singular point (if exists) is a sink for $\alpha > 0$ and a source for $\alpha < 0$. It proves the Corollary in the case $\alpha > 0$ and shows that for $\alpha < 0$ the field (A1, A2) has no limit cycles, homoclinics, and heteroclinics pairs that join the same pair of saddles.

Existence of boundaries $L_+$ for $\delta_1 > 0$ and $L_-$ for $\delta_1 < 0$ for $\alpha < 0$ in the parameter portrait was checked numerically.

### 5.5. Local bifurcations of the vector-fields (A1),(A2) depending on $\alpha$

Let us define in the space of parameters $\{\delta_1, \delta_2, \delta_3, \alpha\}$ of the vector field $J$ the following sets:

$S: \{27a\delta_2^2 = 4\delta_1^3,\ a \neq 0\}$;

$SS: \{\delta_1 = \delta_2 = 0,\ a \neq 0\}$;

$H: \{\delta_3(-3\delta_1 + \delta_3)^2 = 27a\delta_2^2,\ 1 \geq \delta_3 \geq 0\ a \neq 0\}$;

$SH: \{\delta_1 = \delta_3,\ 27a\delta_2^2 = 4\delta_3^2,\ 1 \geq \delta_3 \geq 0,\ a \neq 0\}$;

$CH(\alpha): \{\delta_1 = -\delta_3,\ 27a\delta_2^2 = 16\ \delta_3^3,\ 1 \geq \delta_3 \geq 0,\ L_2(\alpha) < 0,\ a \neq 0,\ \alpha > 0\}$;

It follows from Corollaries 1 and 2 that the singular point $(u_0, 0)$ of vector field (A1, A2) is two-multiple one if the parameters of the vector field $J$ belong to the set $S$ and $u_0 = \pm\sqrt{(\delta_1/3a)}$;

three-multiple one if the parameters belong to the set $SS$ and $u_0 = 0$;

neutral spiral if the parameters belong to the set $H$ and $u_0 = \pm\sqrt{(\delta_3/3a)}$;

neutral spiral with zero first Lyapunov value if the parameters belong to the set $CH(\alpha)$ and $u_0 = \sqrt{(\delta_3/3a)}$;

neutral two-multiple one if the parameters belong to the set *SH* and $u_0=\sqrt{(\delta_3/3a)}$.

Coming back to the initial parameters $k_1$, $k_2$, $\varepsilon$ (see Proposition 2) we get the formulas given in Theorems 1,2.

### 5.6. *End of proofs. Non-local bifurcations of the vector-fields.*

Let us describe boundaries in the parameter space $\{\delta,\alpha\}=\{\delta_1, \delta_2, \delta_3, \alpha\}$ corresponding to non-local bifurcations of vector field *J*. For *positive* $\alpha$ there exist the following bifurcation surfaces (at $\delta_3>0$, see **Fig. 3a**).

The bifurcation *"two-multiple cycles"* is realized on the surface *C*, which touches the surfaces $H^\pm$ by lines $CH^\pm$;

The bifurcation "a *small* loop composed by one of separatrix pairs of the saddle point" is realized on the surfaces $P^+$, $P^-$ and "a *big* loop composed by one of separatrix pairs of the saddle point" is realized on the surfaces $R^+$, $R^-$. Surfaces $S^+$, $H^+$ and $P^+$ have common lines $SH^+$ (see Fig. 3a) as well as $S^-$, $H^-$ and $P^-$ have common lines $SH^-$ (Bogdanov, 1973). Surfaces $R^+$, $R^-$ have common lines of touching with surfaces $S^+$, $S^-$. They also intersect bifurcation surfaces $R^+$, $R^-$ (see Fig. 3) such that four separatrixes compose "8" at the phase plane; this bifurcations was studied by D.Turaev (1985).

For negative $\alpha$ there exist the following bifurcation surfaces (at $\delta_3>0$, see Fig. 5a).

The bifurcation *"a small* loop composed by one of separatrix pairs of the saddle point" is realized on the surfaces $P_+$, $P_-$. Surfaces $S^+$, $H_+$ and $P_+$ have common lines $SH_+$ as well as $S^-$, $H_-$ and $P_-$ have common lines $SH_-$.

The bifurcation "upper, lower (respectively) heteroclinics of saddle singular points" is realized on the surfaces $L_+$, $L_-$ as for $\delta_3>0$ so for $\delta_3<0$. For $\delta_3>0$ $L_+$, $L_-$ and $P_+$, $P_-$ have common line of intersection such that two heteroclinics simultaneously join saddle points.

Generic non-local bifurcations were investigated numerically with the help of program packages [Levitin, 1987, Khibnik et.al, 1993]. In the analysis we used analytical asymptotics of boundaries *C*, $P^\pm$, $P_\pm$ corresponding to non-local bifurcations in vicinities of their touching with boundaries $H^\pm$ of local bifurcations [Bautin, Leontovich, 1976], $S^\pm$ and $H^\pm$ [Bogdanov, 1976] (see also Kuznetsov, 1997, Turaev, 1985).

**Acknowledgements**

The authors thank Carlos Castillo- Chavez for useful discussion of the problem. The work of FB was partially supported by the NSF Grant # HRD0401697.

Fig.1 Neuron spike (in *t-P* plane) from FitzHugh-Nagumo equations. The interpretation of the potential, refractory period, etc. are interpreted from well-known experiments of neuron firing. See e.g., Sherwood (2001).

Fig.2 Phase portrait (in the *P-Q* plane) from FitzHugh-Nagumo equations corresponding to neuron spike.

Fig.3 Schematic parameter portrait of FitzHugh model (3) and the fast wave systems (7+), (13+) of FitzHugh cross-diffusion model (for velocity $C^2 > Dk_1/\varepsilon$ where $D$ is the cross-diffusion coefficient). **a** gives the $\varepsilon-$cut for $0<\varepsilon<1$, and **b** gives the $\varepsilon-$cut for $\varepsilon>1$.

Fig.4. Phase portraits corresponding to parameter domains from Fig.3

Fig.5. Schematic parameter portrait of slow wave systems (7-) of the FitzHugh cross-diffusion model (for the velocity $0<C<\sqrt{(Dk_1/\varepsilon)}$. **a** gives the $\varepsilon-$cut for $0<\varepsilon<1$, and **b** gives the $\varepsilon-$cut for $\varepsilon>1$.

Fig. 6. Phase portraits corresponding to parameter domains from Fig.5

Fig.7. Wave front solutions of a PDE model satisfy boundary conditions: $u \to u_1$ or $u_2$ when $x \to \infty$ or $x \to -\infty$, respectively; they correspond to monotonous *heteroclinic curves* in the $(u,v)$ plane of the ODE wave system. Wave can be a non-monotonous if, for example, one of the phase points is a spiral.

Fig.8. Wave-pulses of a PDE model satisfy boundary conditions: $u \to u_1$ or $u_2$ when $x \to \infty$ and $x \to -\infty$; they correspond to *homoclinic curves* in the $(u,v)$ plane of the ODE wave system. Three typical pictures are given in figures **a**, **b**, and **c**.

Fig.9. Wave-train solutions of a PDE model correspond to *limit cycles* in the $(u,v)$ plane of the ODE wave system.

Fig.10. Phase portraits of model (5) corresponding to three types of standing waves. Figs. a, b, c all possess limit cycles; figs. a, b also have small homoclinics whereas fig c has heteroclinics.

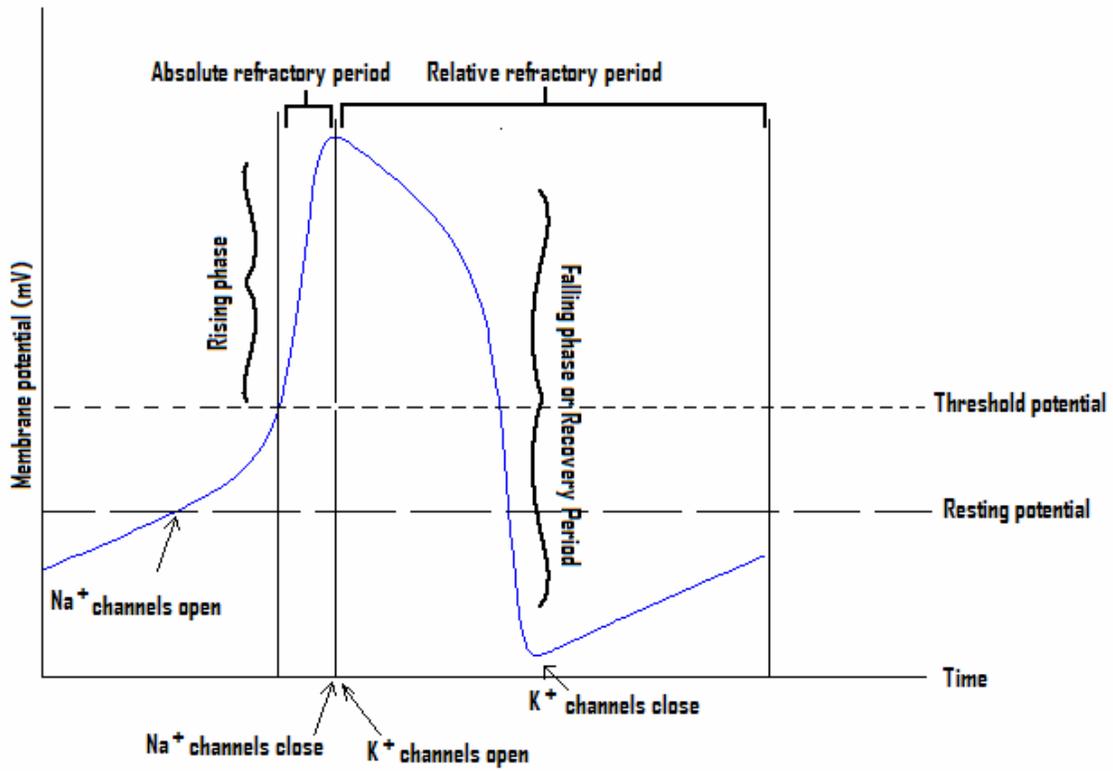

Fig.1 Neuron spike (in *t-P* plane) from FitzHugh-Nagumo equations. The interpretation of the potential, refractory period, etc. are interpreted from well-known experiments of neuron firing. See e.g., Sherwood (2001).

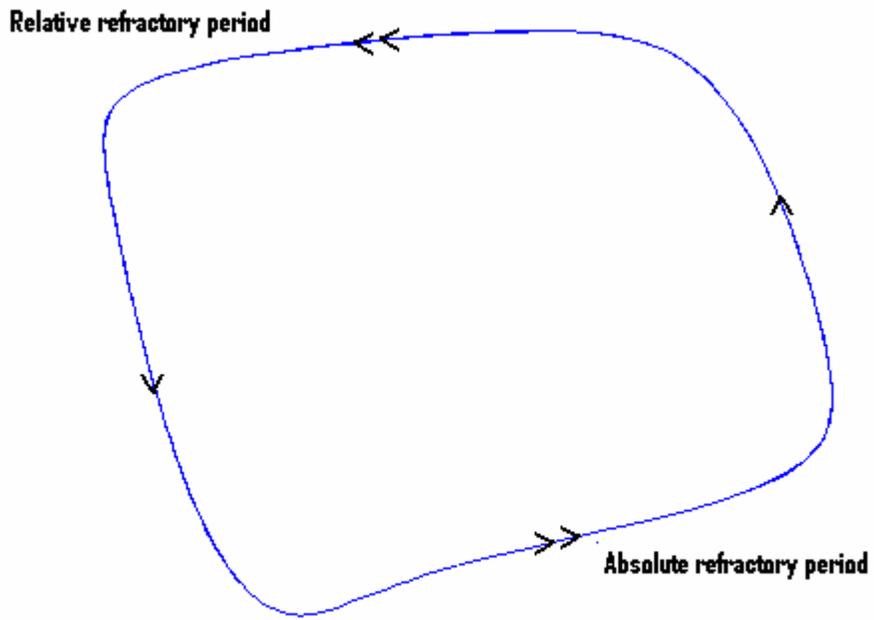

Fig.2 Phase portrait (in the *P-Q* plane) from FitzHugh-Nagumo equations corresponding to neuron spike.

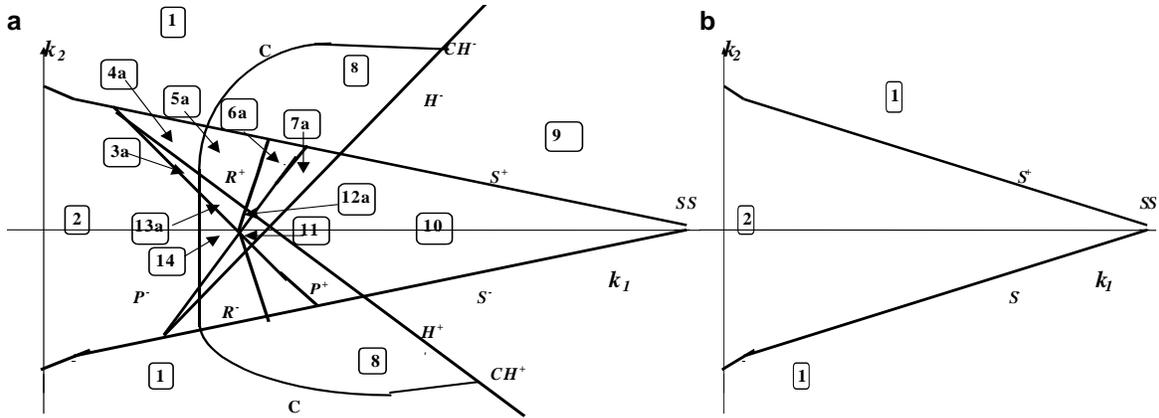

Fig.3 Schematic parameter portrait of FitzHugh model (3) and the fast wave systems (7+), (13+) of FitzHugh cross-diffusion model (for velocity $C^2 > Dk_1/\varepsilon$ where $D$ is the cross-diffusion coefficient).

**a** gives the $\varepsilon$–cut for $0 < \varepsilon < 1$, and **b** gives the $\varepsilon$–cut for $\varepsilon > 1$.

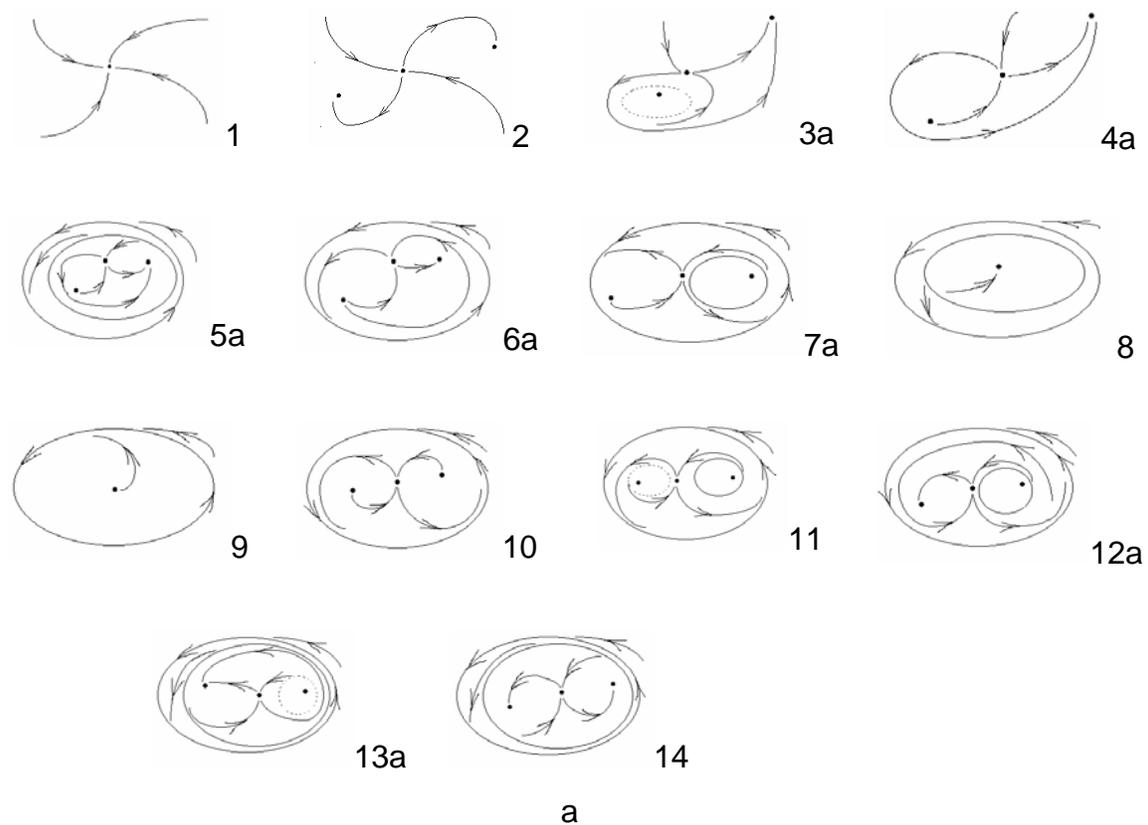

Fig.4. Phase portraits corresponding to parameter domains from Fig.3

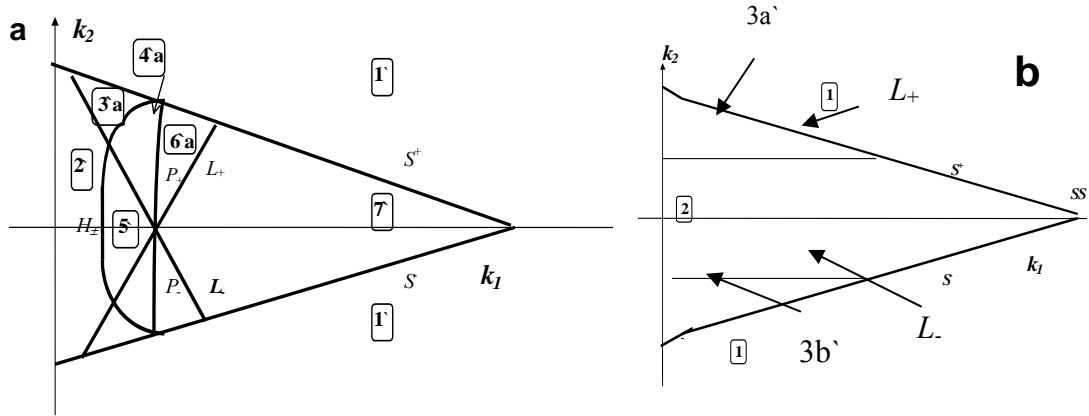

Fig.5. Schematic parameter portrait of slow wave systems (7-) of the FitzHugh cross-diffusion model (for the velocity $0<C<\sqrt{(Dk_1/\varepsilon)}$). **a** is an $\varepsilon$–cut for $0<\varepsilon<1$, **b** is an $\varepsilon$–cut for $\varepsilon>1$.

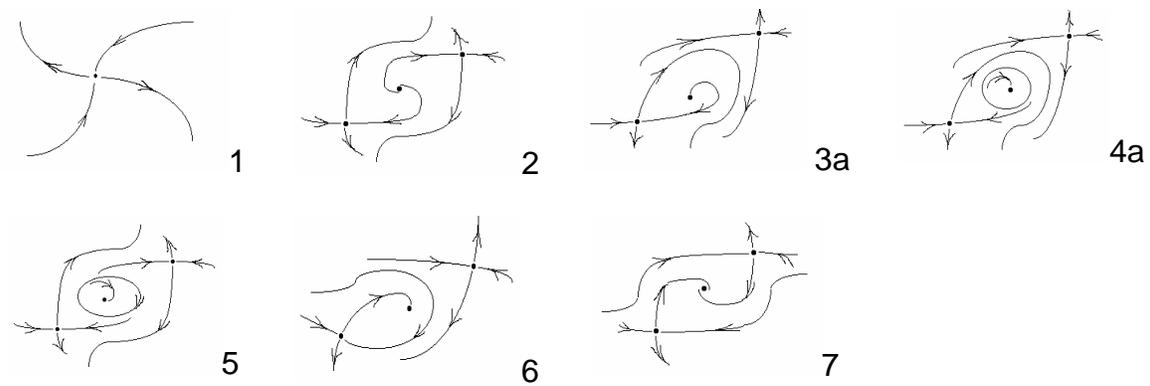

Fig. 6. Phase portraits corresponding to parameter domains from Fig.5

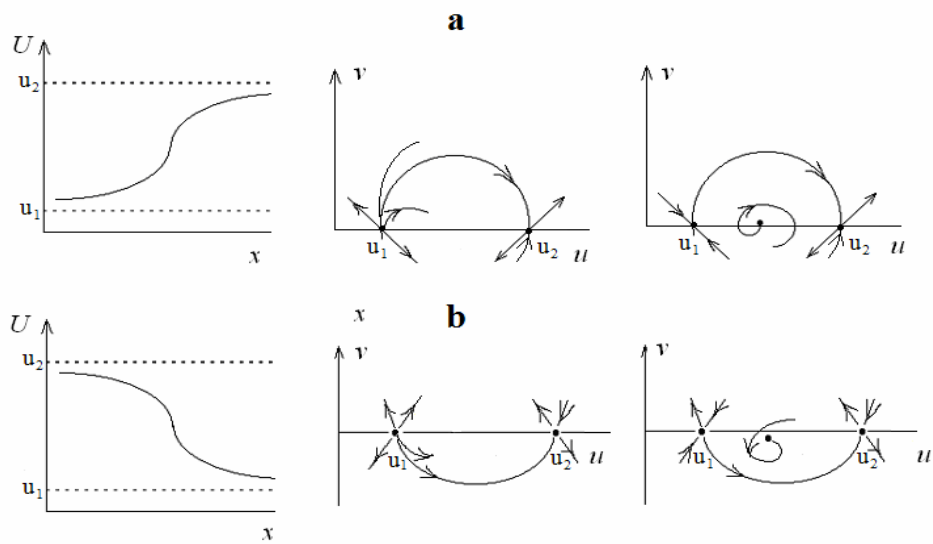

Fig.7. Wave front solutions of a PDE model satisfy boundary conditions: $u \to u_1$ or $u_2$ when $x \to \infty$ or $x \to -\infty$, respectively; they correspond to monotonous *heteroclinic curves* in the $(u,v)$ plane of the ODE wave system. Wave can be a non-monotonous if, for example, one of the phase points is a spiral.

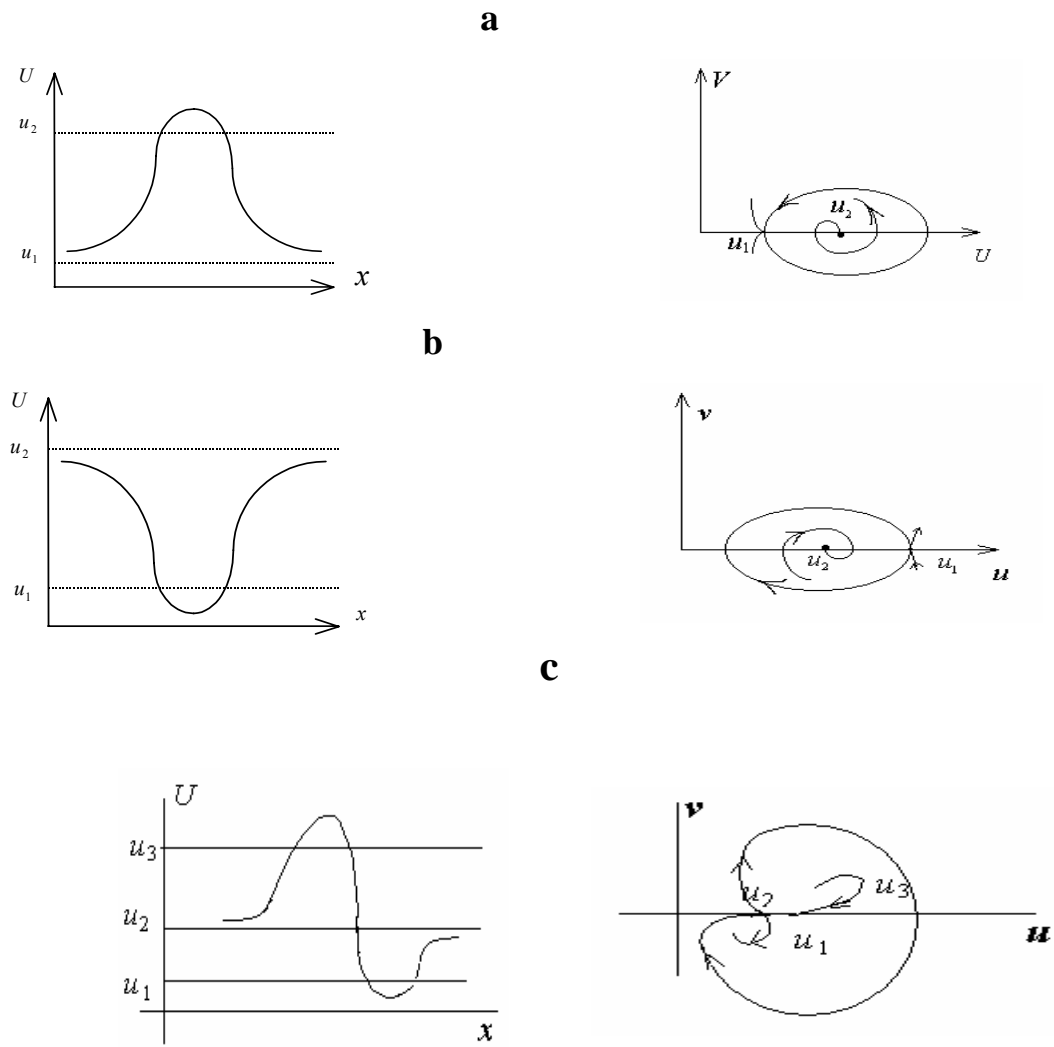

Fig.8. Wave-pulses of a PDE model satisfy boundary conditions: $u \to u_1$ or $u_2$ when $x \to \infty$ and $x \to -\infty$; they correspond to *homoclinic curves* in the $(u,v)$ plane of the ODE wave system. Three typical pictures are given in figures **a**, **b**, and **c**.

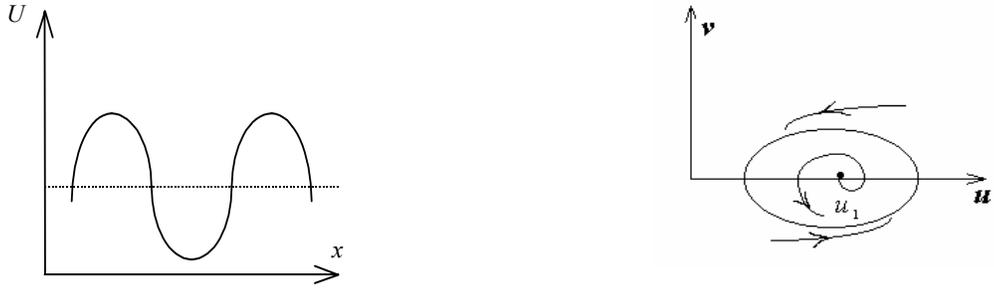

Fig.9. Wave-train solutions of a PDE model correspond to *limit cycles* in the (u,v) plane of the ODE wave system.

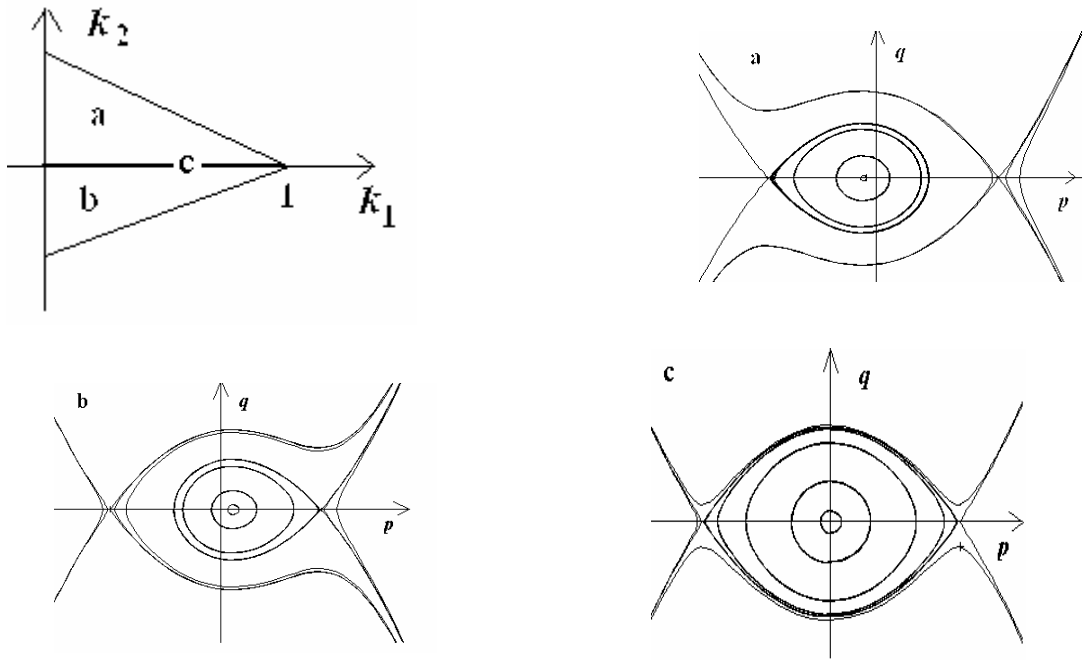

Fig.10. Phase portraits of model (5) corresponding to three types of standing waves. Figs. a, b, c all possess limit cycles; figs. a, b also have small homoclinics whereas fig c has heteroclinics.